\documentclass[twocolumn,conference]{IEEEtran}
\usepackage{amsmath,amssymb}
\usepackage[dvips]{graphicx}
\usepackage{subfigure}
\usepackage{algorithm,algorithmic}
\usepackage{booktabs}
\usepackage{multicol}
\usepackage{multirow}
\usepackage{cite,color,bm}
\usepackage{amsthm}
\usepackage{epsfig}
\usepackage{epstopdf}
\usepackage{subfigure}

\ifCLASSINFOpdf

\else

\fi

\begin{document}
\title{Sparse Array Beampattern Synthesis via Majorization-Based ADMM}
\author{
\IEEEauthorblockN{Tong~Wei, Linlong~Wu, and Bhavani~Shankar~M.~R.\\}
\IEEEauthorblockA{Interdisciplinary Centre for Security, Reliability and Trust (SnT), University of Luxembourg\\
Email: \{tong.wei, linlong.wu, bhavani.shankar\}@uni.lu}
}

\maketitle

\begin{abstract}
Beampattern synthesis is a key problem in many wireless applications. With the increasing scale of MIMO antenna array, it is highly desired to conduct beampattern synthesis on a sparse array to reduce the power and hardware cost. In this paper, we consider conducting beampattern synthesis and sparse array construction jointly. In the formulated problem, the beampattern synthesis is designed by minimizing the matching error to the beampattern template, and the Shannon entropy function is first introduced to impose the sparsity of the array. Then, for this nonconvex problem, an iterative method is proposed by leveraging on the alternating direction multiplier method (ADMM) and the majorization minimization (MM). Simulation results demonstrate that, compared with the benchmark, our approach achieves a good trade-off between array sparsity and beampattern matching error with less runtime.
\end{abstract}
\IEEEpeerreviewmaketitle


\section{Introduction}
\label{sec:intro}
Beampattern synthesis aims to design the appropriate weight vector to achieve a desired radiation pattern. It has been and continues to be a widely researched topic in  radar and wireless communication systems \cite{01,02,03}. Especially for the latter, the beamforming technique expresses  several advantages, including improved signal to interference and noise
ratio (SINR), reduced interference and enhanced security \cite{04}.

Recently, sparse array structures have attracted the significant research interest due to their inherent capability in source localization, simplified feeding networks and the reduced hardware  cost and power consumption\cite{05,06,07}. With regards to beampattern synthesis, the sparse array configuration with the minimum number of elements is also required to achieve a specified performance \cite{08,09}. Specifically, we  aim to design a desired beampattern by selecting only a few elements from a predefined array.  On the one hand, all the antenna resources need to be exploited to flexibly achieve different beampatterns; on the other hand, the sparse configuration is expected to reduce the overall system cost and power consumption. Hence,  beampattern synthesis and antenna selection should be considered simultaneously in order to achieve an appropriate trade-off between these two requirements.

In general, the problem of sparse array beampattern design usually formulated as $\ell_0$-norm minimization problem with a predefined pattern shape constraint \cite{08,10,11,12}. The central problem in these works is the formulation of different methodologies to solve the $\ell_0$-norm optimization problem. In \cite{08}, the weighted $\ell_1$-norm is first utilized to approximate the nonconvex $\ell_0$-norm. Then, an iterative method is introduced to solve the second-order cone programming (SOCP) problem. To further improve the sparsity of the solution, the $\ell_p$-norm regularization, where $0<p<1$,  is used in \cite{10}, and then the alternating direction multiplier method (ADMM) framework is directly utilized to deal with the nonconvex optimization problem. Recently, the optimal selection vector (or matrix) is introduced in beampattern design when the prior information (i.e. sparsity level) is provided \cite{13,14,15}. However, all the aforementioned methods cannot  balance the beampattern design and the sparsity of array configuration simultaneously.

In this paper, we consider the problem of sparse array beampattern synthesis. Compared with the existing weighted $\ell_{1}$-norm algorithms \cite{07,08}, the Shannon entropy regularization, which can simultaneously improve the sparsity level and increase the value of nonzero weight, is first utilized to better prompt the sparsity of array configuration. Meanwhile, different from \cite{13,14,15}, the proposed method does not require to predefine the sparsity level. The resulting nonconvex problem is effectively solved by the majorization-based ADMM, which combines the majorization minimization (MM) and the ADMM. Numerical results demonstrate the effectiveness of the proposed method, both in terms of convergence and balance between sparsity and beamshaping.

\emph{Notations:} $(\cdot)^T$, $(\cdot)^{\ast}$ and $(\cdot)^H$ denote transpose, conjugate, and Hermitian transpose, respectively. $\Re(\cdot)$ denotes the real part of a complex value. ${\bf 1}$ and ${\bf I}$  denote  all one vector and the identity matrix, respectively. $\|\cdot\|_p$ denotes $\ell_{p}$-norm.

\section{Problem Formulation}
Let us consider a transmit array with $N$ isotropic antennas uniformly placed with the inter-element spacings $d$. Then, the corresponding transmit steering vector is
\begin{equation}\label{1}
  {\bf a}(\theta)=[1,e^{j\frac{2\pi}{\nu}d\sin(\theta)},\cdots,e^{j\frac{2\pi}{\nu}(N-1)d\sin(\theta)}]^T\in\mathbb{C}^{N\times1},
\end{equation}
where $\theta$ belongs to the whole angle space $\Theta\triangleq[-90^{\circ},+90^{\circ}]$ and $\nu$ denotes the wavelength. The transmit beampattern is given by
\begin{equation}\label{2}
  P(\theta)={\bf w}^H{\bf A}(\theta){\bf w},
\end{equation}
where ${\bf A}(\theta)={\bf a}(\theta){\bf a}^H(\theta)$ and ${\bf w}=[w_1, w_2,\cdots,w_N]^T\in\mathbb{C}^{N\times1}$ denotes the weight vector. Without loss of generality, we set $\|{\bf w}\|_2^2=1$, which means that the array operates in the maximal power model.

In practice, especially for a large-scale array, it is desired to reduce the hardware cost and power consumption, which can be achieved  by deploying a sparse array. Mathematically, sparsity regularization will be imposed into the beampattern design formulation, in which the array sparsity and the beampattern should be well balanced. In light of this trade-off, the problem of interested is formulated as
\begin{equation}\label{3}
  \begin{split}
     \min_{\alpha,{\bf w}} & {\quad} \lambda\sum_{k=1}^{K}\|{\bf w}^H{\bf A}(\theta_k){\bf w}-\alpha d(\theta_k)\|_2^2+f({\bf w})  \\
               s.t.        & {\quad} \|{\bf w}\|_2^2=1,
   \end{split}
\end{equation}
where $\lambda$ is the trade-off parameter, $f({\bf w})$ denotes the sparsity-promoting regularization function, $\theta_k$ denotes the $k$-th angle within the angle space $\Theta$ and $\alpha$ is used to scale the desired beampattern $d(\theta_k)$.

\begin{figure}[tbh]
\begin{centering}
\includegraphics[scale=0.225]{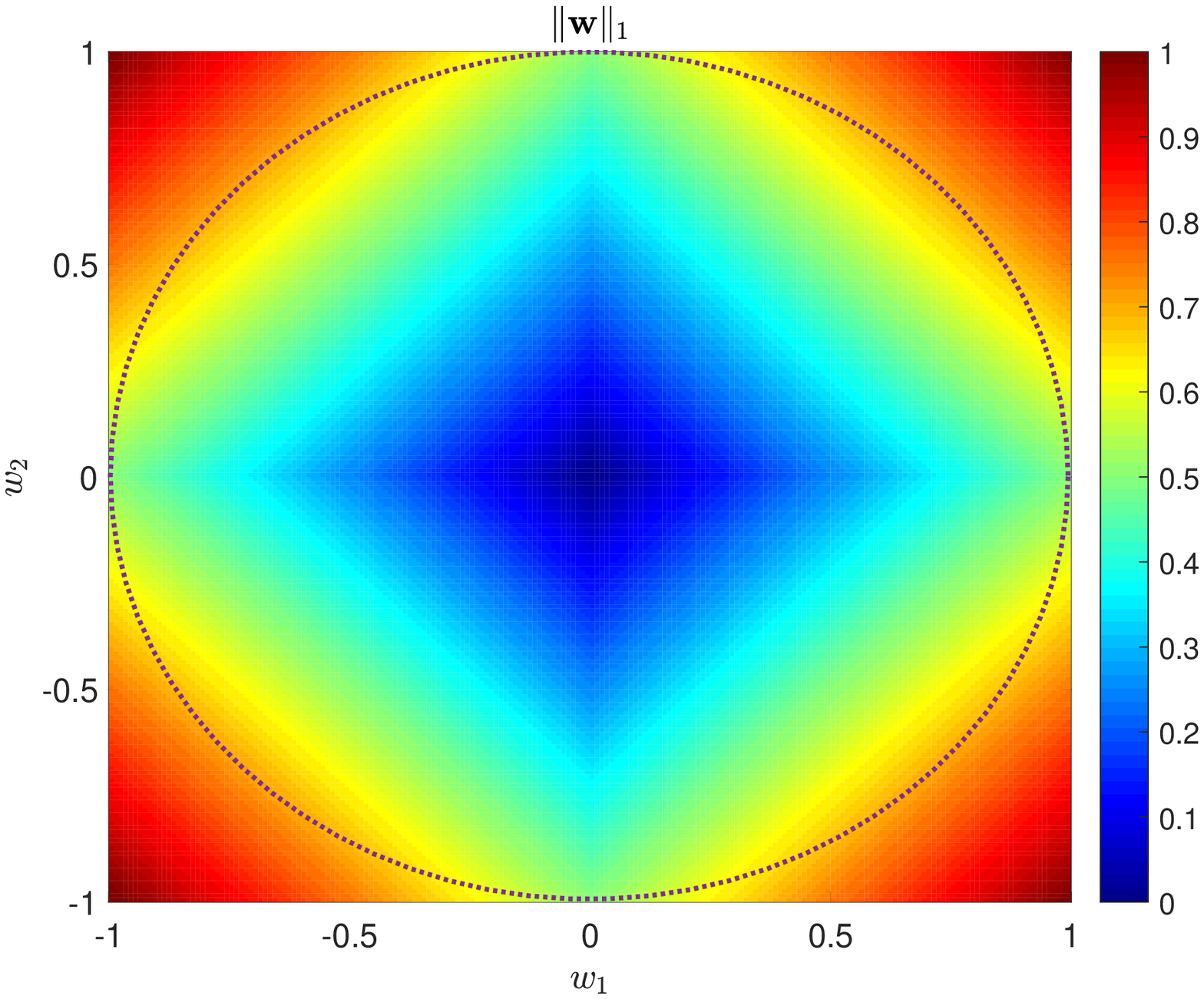}\includegraphics[scale=0.225]{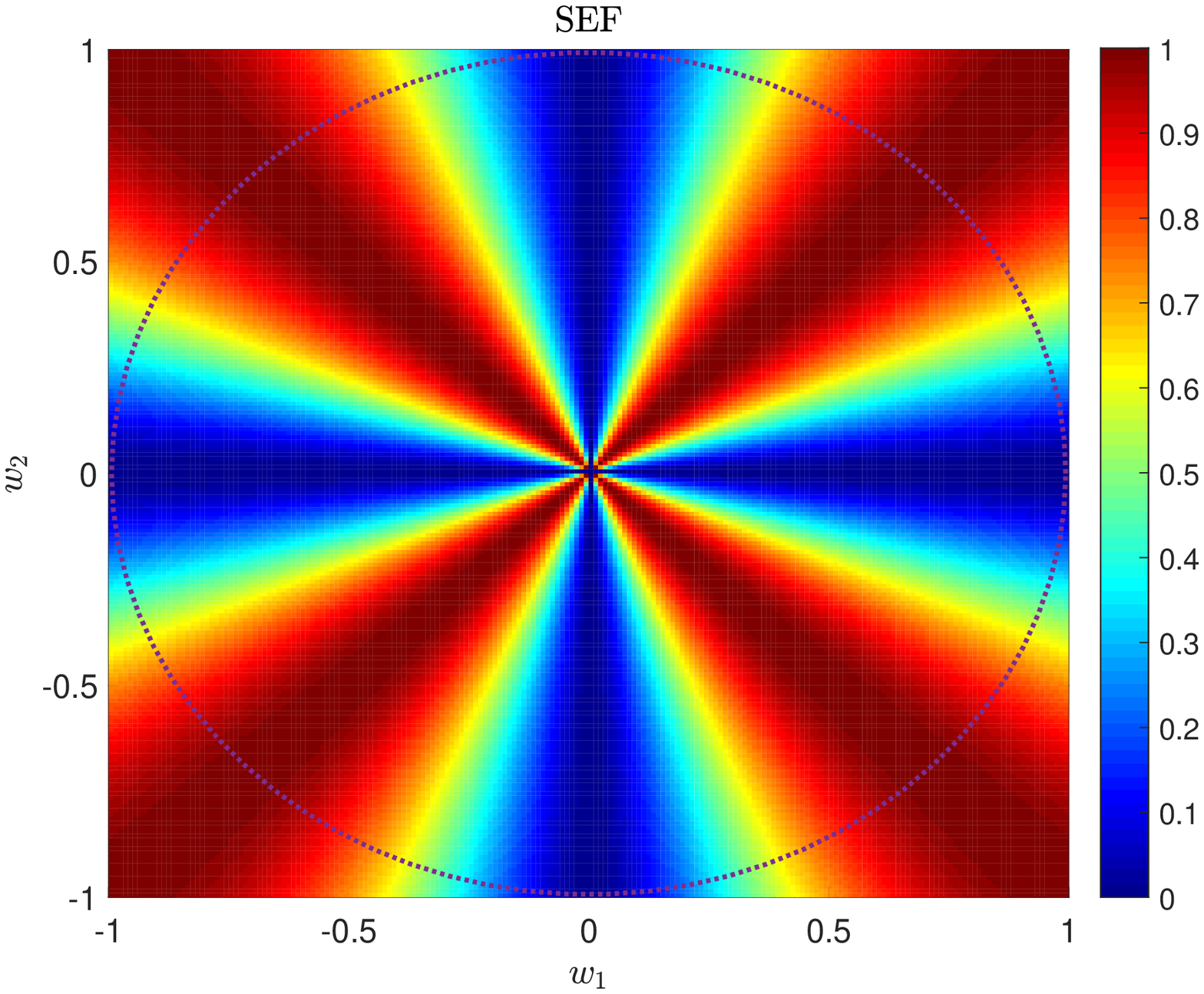}
\par\end{centering}
\caption{Comparison of different sparsity-promoting regularization functions.}
\label{fig:picture001}
\end{figure}

It is noted that there exist various choices to promote the sparse solution of problem \eqref{3}. For example, the well-known $\ell_1$-norm is utilized to achieve the sparsity solution \cite{07,08}. Recently, the Shannon entropy function (SEF) is being widely used in compressed sensing \cite{16,17}, due to its ability to measure the concentration and diversity of a vector. Its definition in terms of $\bf{w}$ is given by
\begin{equation}\label{4}
  f\left(\bf{w}\right)=\sum_{n=1}^{N}\left(\frac{|w_n|^2}{\|{\bf w}\|_2^2}\right)\log\frac{|w_n|^2}{\|{\bf w}\|_2^2}.
\end{equation}
Fig.1 compares the commonly used $\ell_l$-norm with the SEF. It is seen that both of them can prompt sparsity. However, the SEF can simultaneously improve the sparsity level and increase the value of nonzero entries of ${\bf w}$. Especially on unit sphere, i.e., $\|{\bf w}\|_2^2=1$, the SEF has better sparsity-promoting ability compared with $\ell_1$-norm. Due to the superior properties of the SEF, in this paper, we consider it as the sparsity-promoting regularizer in problem \eqref{3}.

\section{Majorization-Based ADMM Optimization Algorithm}
\label{sec:pagestyle}
Note that problem \eqref{3} is still nonconvex due to both the objective function and the constraint. Hence, we will derive an iterative algorithm based on the powerful MM and ADMM frameworks. To tackle the nonconvex quartic objective, we introduce an auxiliary variable ${\bf v}\in\mathbb{C}^{N\times1}$ to convert problem \eqref{3} into
\begin{equation}\label{5}
  \begin{split}
   \min_{\alpha,{\bf w},{\bf v}} & {\quad} \lambda\varphi(\alpha,{\bf v},{\bf w})+f({\bf w})\\
             s.t. & {\quad} {\bf w}-{\bf v} = {\bf 0} \\
                  & {\quad} \|{\bf w}\|_2^2=1,
   \end{split}
\end{equation}
where $\varphi(\alpha,{\bf v},{\bf w})=\sum_{k=1}^{K}\|{\bf w}^H{\bf A}(\theta_k){\bf v}-\alpha d(\theta_k)\|_2^2$.
Then, the augmented Lagrangian of problem \eqref{5} is
\begin{equation}\label{6}
  \begin{split}
  \mathcal{L}(\alpha,{\bf v},{\bf w},{\bf u})&=\lambda\sum_{k=1}^{K}\|{\bf w}^H{\bf
             A}(\theta_k){\bf v}-\alpha d(\theta_k)\|_2^2+f({\bf w})\\
             &\quad +{\bm \gamma}^H({\bf w}-{\bf v})+\frac{\rho}{2}\|{\bf w}-{\bf v}\|_2^2 \\
             &=\lambda\sum_{k=1}^{K}\|{\bf w}^H{\bf A}(\theta_k){\bf v}-\alpha d(\theta_k)\|_2^2+f({\bf w})\\
             &\quad +\frac{\rho}{2}\|{\bf w}-{\bf v}+{\bf u}\|_2^2+const.
   \end{split}
\end{equation}
where ${\bf u}=\frac{1}{\rho}{\bm \gamma}$ denotes the dual variable \cite{18} and $\rho$ is a positive penalty parameter. Within the framework of ADMM, the update rules at the $(t+1)$-th iteration are given by
\begin{subequations} \label{7}
  \begin{align}
  {\alpha}^{(t+1)}&:=\mathop{\mathrm{arg min}}\limits_{\alpha} \ \mathcal{L}(\alpha,{\bf v}^{(t)},{\bf w}^{(t)},{\bf u}^{(t)}),  \\
  {\bf v}^{(t+1)}&:=\mathop{\mathrm{arg min}}\limits_{{\bf v}} \ \mathcal{L}(\alpha^{(t+1)},{\bf v},{\bf w}^{(t)},{\bf u}^{(t)}),  \\
  {\bf w}^{(t+1)}&:=\mathop{\mathrm{arg min}}\limits_{\|{\bf w}\|_2^2=1} \ \mathcal{L}(\alpha^{(t+1)},{\bf v}^{(t+1)},{\bf w},{\bf u}^{(t)}),   \\
  {\bf u}^{(t+1)}&:={\bf u}^{(t)}+({\bf w}^{(t+1)}-{\bf v}^{(t+1)}).
  \end{align}
\end{subequations}

\subsection{Update of $\alpha$}
At the $(t+1)$-th iteration, given ${\bf v}^{(t)}$ and ${\bf w}^{(t)}$, the optimization problem (7a) can be written as
\begin{equation}\label{8}
 \min_{\alpha} {\quad} \sum_{k=1}^{K}\|{{\bf w}^{(t)}}^H{\bf A}(\theta_k){\bf v}^{(t)}-\alpha d(\theta_k)\|_2^2,
\end{equation}
which has the closed-form solution
\begin{equation}\label{9}
  {\alpha}^{(t+1)}=\frac{{\bm\Xi}_1^{(t)}}{\sum_{k=1}^{K}d^2(\theta_k)},
\end{equation}
with ${\bm\Xi}_1^{(t)}=\sum_{k=1}^{K}2d(\theta_k){\Re}\left({{\bf w}^{(t)}}^H{\bf A}(\theta_k){\bf v}^{(t)}\right)$.

\subsection{Update of ${\bf v}$}

At the $(t+1)$-th iteration, given ${\alpha}^{(t+1)}$, ${\bf w}^{(t)}$ and ${\bf u}^{(t)}$, we can update ${\bf v}$ by solving the following problem
\begin{equation}\label{10}
   \min_{\bf v} {\quad} \lambda\varphi(\alpha^{(t+1)},{\bf v},{\bf w}^{(t)}) +\frac{\rho}{2}\|{\bf w}^{(t)}-{\bf v}+{\bf u}^{(t)}\|_2^2.
\end{equation}
Noticed that problem \eqref{10} is an unconstrained quadratic programming problem and hence convex.
By setting the derivative of objective function of \eqref{10} with respect to ${\bf v}^{\ast}$ to be zero, we have
\begin{equation}\label{11}
  {\bm\Xi}_2^{(t)}{\bf v}-{\bm\Upsilon}_2^{(t+1)}{\bf w}^{(t)}+\frac{\rho}{2}\left({\bf v}-({\bf w}^{(t)}+{\bf u}^{(t)})\right)={\bf 0}
\end{equation}
where
\begin{subequations} \label{12}
  \begin{align}
  &{\bm\Xi}_2^{(t)} = \lambda\sum_{k=1}^{K}{\bf A}^H(\theta_k){{\bf w}^{(t)}}{{\bf w}^{(t)}}^H{\bf A}(\theta_k),   \\
  &{\bm\Upsilon}_2^{(t+1)} = \lambda\sum_{k=1}^{K}\alpha^{(t+1)} d(\theta_k){\bf A}^H(\theta_k).
\end{align}
\end{subequations}
The  solution to \eqref{11} is
\begin{equation}\label{13}
  {\bf v}^{(t+1)}=({\bm\Xi}_2^{(t)}+\frac{\rho}{2}{\bf I})^{-1}({\bm\Upsilon}_2^{(t+1)}{\bf w}^{(t)}+\frac{\rho}{2}({\bf w}^{(t)}+{\bf u}^{(t)})),
\end{equation}
which is also the optimal solution to problem \eqref{10}.

\subsection{Update of ${\bf w}$}
At the $(t+1)$-th iteration, given ${\alpha}^{(t+1)}$, ${\bf v}^{(t+1)}$, ${\bf w}^{(t)}$  and ${\bf u}^{(t)}$, we can update ${\bf w}$ into solving the following problem
\begin{equation}\label{14}
  \begin{split}
   \min_{\bf w} & \quad \lambda\varphi(\alpha^{(t+1)},{\bf v}^{(t+1)},{\bf w})+f({\bf w})  \\
                 &\qquad \qquad +\frac{\rho}{2}\|{\bf w}-{\bf v}^{(t+1)}+{\bf u}^{(t)}\|_2^2 \\
    s.t. & {\quad} \|{\bf w}\|_2^2=1.
    \end{split}
\end{equation}
Due to the concave nature of $f({\bf w})$, it is difficult to directly solve problem \eqref{14}. Thus, we resort to the majorization-minimization (MM) framework \cite{19,20}, and solving the problem \eqref{14} is then transformed to solving a series of subproblems until convergence.

To begin with, let us introduce an important majorizing function of $f(\bf w)$ via the following Lemma.

\newtheorem{lemma}{Lemma}
\begin{lemma} \label{lemma}
For any complex set ${\bf w}\in\mathbb{C}^{N\times1}$ with $\|{\bf w}\|_2^2=1$, we always have
\begin{equation}\label{15}
  f({\bf w})=g(\widetilde{\bf w})\leq{\bf w}^H{\bf D}_t{\bf w}+const.
\end{equation}
where $\widetilde{\bf w}\!=\!{\bf w}\odot{\bf w}^{\ast}$, $g(\widetilde{\bf w})\!=\!-\sum_{n=1}^{N}\widetilde{w}_n \log \widetilde{w}_n$, $const.\!=\! g(\widetilde{\bf w}^{(t)})-{\nabla g(\widetilde{\bf w}^{(t)})}^T\widetilde{\bf w}^{(t)}$, ${\bf D}_t\!=\!\mathrm{diag}(\nabla g(\widetilde{\bf w}^{(t)}))$ and $\nabla g(\widetilde{\bf w}^{(t)})$ is the gradient vector with the $n$-th element $\nabla g(\widetilde{w}_n^{(t)})\!=\! -\log\widetilde{w}_n^{(t)}-1, n\!=\!1,\cdots,N$.
\end{lemma}

\begin{IEEEproof}
Recalling the constraint $\|{\bf w}\|_2^2=\!1$, we have
\begin{equation}\label{16}
  f({\bf w})=-\sum_{n=1}^{N}w_n w_n^{\ast}\log w_n w_n^{\ast}.
\end{equation}
Hence, it is easily derived that $f({\bf w})\!=\!g(\widetilde{\bf w})$. Then, the upper bound function of $g(\widetilde{\bf w})$ at current point $\widetilde{\bf w}^{(t)}$  is
\begin{equation}\label{17}
  \begin{split}
  g(\widetilde{\bf w}) & \leq g(\widetilde{\bf w}^{(t)})+\nabla g(\widetilde{\bf w}^{(t)})^T(\widetilde{\bf w}-\widetilde{\bf w}^{(t)}) \\
                       & =\nabla g(\widetilde{\bf w}^{(t)})^T({\bf w}\odot{\bf w}^{\ast})+const. \\
                       & = {\bf w}^{H}{\bf D}_t{\bf w}+const.
   \end{split}
\end{equation}
where $const.\!=\! g(\widetilde{\bf w}^{(t)})-{\nabla g(\widetilde{\bf w}^{(t)})}^T\widetilde{\bf w}^{(t)}$ and $\nabla g(\widetilde{\bf w}^{(t)})$ stands for the derivative of $g(\widetilde{\bf w}^{(t)})$ with respect $\widetilde{\bf w}^{(t)}$, whose $n$-th entry is
\begin{equation}\label{18}
  \nabla g(\widetilde{w}_n^{(t)}) = -\log\widetilde{w}_n^{(t)}-1, n=1,\cdots,N.
\end{equation}
Based on above, we conclude that  \eqref{15} is satisfied, thereby completing the proof.
\end{IEEEproof}

Hence, replacing the function $f({\bf w})$ by its majorizer from \eqref{15} and ignoring the constants, problem \eqref{14} can be simplified as follows
\begin{equation}\label{19}
  \begin{split}
     \min_{{\bf w}} &{\quad} \lambda\varphi(\alpha^{(t+1)},{\bf v}^{(t+1)},{\bf w})+{\bf w}^H{\bf D}_t{\bf w}\\
                    & \qquad \qquad +\frac{\rho}{2}\|{\bf w}-{\bf v}^{(t+1)}+{\bf u}^{(t)}\|_2^2\\
           s.t.     & {\quad} \|{\bf w}\|_2^2=1.
   \end{split}
\end{equation}

Noted that problem \eqref{19} is difficult to obtain the global optimal solution due to the nonconvex unit sphere constraint. Even though this kind of problem can be solved by semidefinite relaxation (SDR) \cite{21}, the corresponding problem size will greatly grow which leads to higher computational complexity. Herein, we utilize a more effective method, which named projected gradient descent (PGD) \cite{22}, to tackle problem \eqref{19}. Specifically, we can first remove the unit sphere constraint and solve the unconstrained problem. Then, the projection operator is utilized to project the solution onto unit sphere.

Setting the derivative of objective function of \eqref{19} with respect to ${\bf w}^{\ast}$ as zero, we have
\begin{equation}\label{20}
  {\bm\Xi}_3^{(t+1)}{\bf w}-{\bm\Upsilon}_3^{(t+1)}{\bf v}^{(t+1)}+{\bf D}_t{\bf w}+\frac{\rho}{2}({\bf w}-({\bf v}^{(t+1)}-{\bf u}^{(t)}))\!=\!{\bf 0}
\end{equation}
where
\begin{subequations} \label{21}
  \begin{align}
  &{\bm\Xi}_3^{(t+1)} = \lambda\sum_{k=1}^{K}{\bf A}(\theta_k){{\bf v}^{(t+1)}}{{\bf v}^{(t+1)}}^H{\bf A}^H(\theta_k),   \\
  &{\bm\Upsilon}_3^{(t+1)} = \lambda\sum_{k=1}^{K}\alpha^{(t+1)} d(\theta_k){\bf A}(\theta_k).
\end{align}
\end{subequations}
According to \eqref{20},  it is concluded that
\begin{small}
\begin{equation}\label{22}
  \widehat{{\bf w}}^{(t+1)}\!=\!({\bm\Xi}_3^{(t+1)}+{\bf D}_t+\frac{\rho}{2}{\bf I})^{-1}({\bm\Upsilon}_3^{(t+1)}{\bf v}^{(t+1)}+\frac{\rho}{2}({\bf v}^{(t+1)}-{\bf u}^{(t)})).
\end{equation}
\end{small}

\begin{algorithm}[!t]
  \caption{Majorization-based ADMM for Problem \eqref{3}}
  \label{algorithm1}
  \hspace*{0.02in} {\bf Input:}
   $N, \lambda, \rho, \eta, {\bf A}(\theta_k), d(\theta_k), k=1,\cdots,K$ \\
  \hspace*{0.02in} {\bf Initalize:}
   $\alpha^{(0)}, {\bf v}^{(0)}, {\bf w}^{(0)}$, ${\bf u}^{(0)}$ and counter $t\!=\!0$

  \begin{algorithmic}[1]
    \REPEAT
      \STATE Update $\alpha^{(t+1)}$ using \eqref{9}
      \STATE Update ${\bf v}^{(t+1)}$ using \eqref{13}
      \STATE Calculate gradient vector $\nabla g(\widetilde{\bf w}^{(t)})$ via \eqref{18}
      \STATE Reconstruct the diagonal matrix ${\bf D}_t$ using $\nabla g(\widetilde{\bf w}^{(t)})$
      \STATE Update ${\bf w}^{(t+1)}$ using \eqref{23}
      \STATE Update ${\bf u}^{(t+1)}$ using \eqref{24}
      \STATE Counter Increase: $t\!\leftarrow \!t\!+\!1$
    \UNTIL{$\|{\bf w}^{(t+1)}-{\bf w}^{(t)}\|_2\leq \eta$}
  \end{algorithmic}
  \hspace*{0.02in} {\bf Output:}
  ${\bf w}^{\star}={\bf w}^{(t)}$
\end{algorithm}
The solution to \eqref{19} is
\begin{equation}\label{23}
  {\bf w}^{(t+1)}=\mathcal{P}(\widehat{{\bf w}}^{(t+1)})
\end{equation}
where $\mathcal{P}(\cdot)=(\cdot)/\|\cdot\|_2$ denotes the spherical projection.

\subsection{Update of ${\bf u}$}
At the $(t+1)$-th iteration, given ${\bf v}^{(t+1)}$, ${\bf w}^{(t+1)}$ and ${\bf u}^{(t)}$, the dual variable can be directly updated as
\begin{equation}\label{24}
  {\bf u}^{(t+1)}={\bf u}^{(t)}+({\bf w}^{(t+1)}-{\bf v}^{(t+1)}).
\end{equation}

According to above discussions, the proposed majorization-based ADMM based algorithm for solving problem \eqref{3} is summarized in \textbf{Algorithm 1}.

\section{Simulation Results}
In this section, some representative numerical examples are provided to evaluate the performance of proposed method for sparse array transmit beampattern synthesis. Herein, the spatial domain $\Theta\!\triangleq\![-90^{\circ},+90^{\circ}]$ is uniformly sampled with step-size $1^{\circ}$. Further, the initial value of ${\bf v}^{(0)}$ and ${\bf w}^{(0)}$ are randomly generated from zero-mean complex Gaussian distribution and normalized to $\|{\bf v}^{(0)}\|_2^2=1$ and $\|{\bf w}^{(0)}\|_2^2=1$. Meanwhile, we set $\alpha^{(0)}=1$ and ${\bf u}^{(0)}={\bf 0}$. Throughout the simulations, other parameters are set as $N=30$, $\lambda=0.1$,  $\rho=30$ and $\eta=10^{-8}$. Finally $\hat{N}$ denotes the number of selected antennas.

\subsection{Beampattern with a single mainlobe}
\begin{figure}[!t]
\centering
\includegraphics[width=1\linewidth]{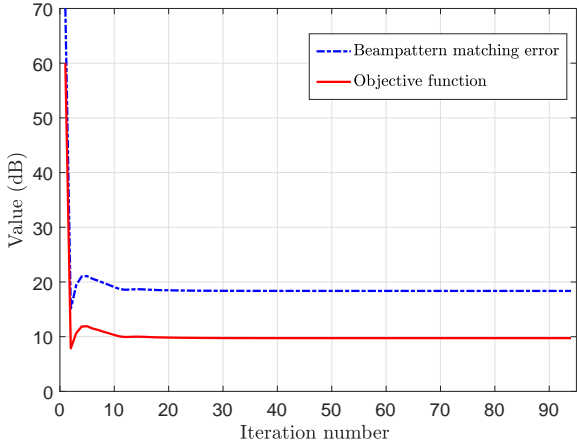}
\caption{Convergence of the objective and beampattern matching; selected number of elements, $\hat{N}=18$.}
\label{Fig1}
\end{figure}

\begin{figure}[!t]
\centering
\includegraphics[width=1\linewidth]{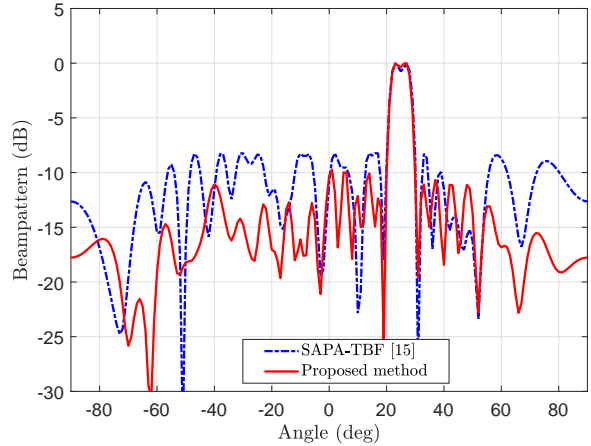}
\caption{Beampattern comparison, $\hat{N}=18$.}
\label{Fig2}
\end{figure}

In this example, we consider a  mainlobe  spanning $\Theta_m=[22^{\circ},28^{\circ}]$ and let the desired pattern as $d(\theta_m)=1000,\theta_m\in\Theta_m$. Fig.2 demonstrates the convergence performance of proposed method. It is seen that the proposed method expresses good convergence performance for both objective value and beampattern matching within 100 iterations. Fig.3 compares the synthesized beampatterns. The benchmark method, named SAPA-TBF \cite{15}, needs to predefine the sparsity level first and then minimize the maximal difference between the designed beamapttern and the desired one. For comparison, we set $\hat{N}=18$ for SAPA-TBF, which is the number of selected elements of our proposed method. It is seen from Fig.3 that the proposed method has relatively lower sidelobes than SAPA-TBF. And the normalized beampattern matching error of proposed method is $-1.651$ dB compared with $2.628$ dB for SAPA-TBF method. Meanwhile, in \cite{15}, it is proved that SAPA-TBF has better beampattern synthesis performance than the $\ell_p$-norm based method \cite{10}.

\subsection{Beampattern with two mainlobes}

\begin{figure}[!t]
\centering
\includegraphics[width=1\linewidth]{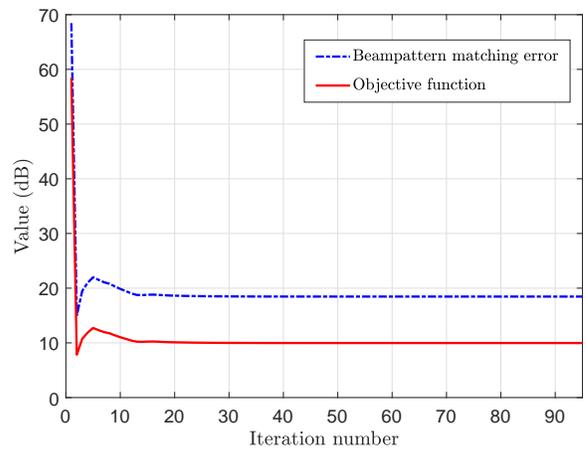}
\caption{Convergence of the objective and beampattern matching; selected number of elements,  $\hat{N}=20$.}
\label{Fig3}
\end{figure}

\begin{figure}[!t]
\centering
\includegraphics[width=1\linewidth]{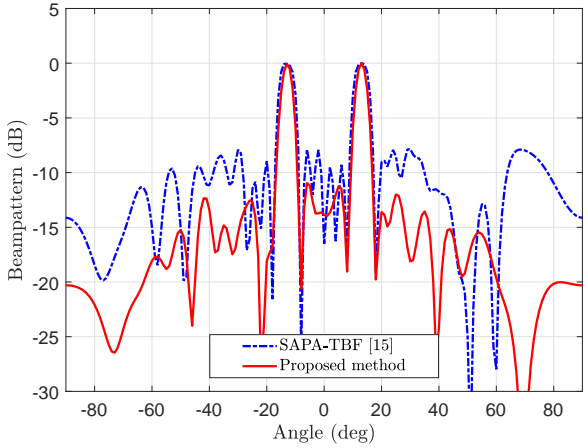}
\caption{Beampattern comparison , $\hat{N}=20$.}
\label{Fig4}
\end{figure}

In this example, we consider the two mainlobes which are located in
$\Theta_m\!=\![-15^{\circ},-11^{\circ}]\cup[11^{\circ},15^{\circ}]$. Meanwhile, the desired beampattern is set as  $d(\theta_m)=1000,\theta_m\in\Theta_m$. We set $\hat{N}=20$ for SAPA-TBF. Fig.4 expresses the convergence performance of proposed method. Again, the proposed method still has good convergence performance within 100 iterations. Fig.5 compares
the synthesized beampatterns. It is observed that our method has a lower peak side level than SAPA-TBF. The corresponding  normalized beampattern matching error for our method is $0.869$ dB, while $3.838$ dB in \cite{15}, indicating improved performance of our method in beampattern matching.

Table 1. compares the cardinality, required runtime and normalized beampattern matching error of these two methods. It is seen that, with the same selected antennas, the proposed method has the smaller beampattern matching error compared with SAPA-TBF. Meanwhile, our method is more efficiency than SAPA-TBF in terms of runtime.

\label{sec:majhead}
\begin{table}
  \centering
  \caption{The performance comparison of different methods.}
  \vspace{0.1cm}
  \label{tab:performance_comparison}
    \begin{tabular}{|c|c|c|c|c|}
    \hline
    \multirow{2}{*}{Method}&
    \multicolumn{2}{c|}{Single mainlobe}&\multicolumn{2}{c|}{Two mainlobes}\cr\cline{2-5}
    & Proposed &SAPA-TBF &Proposed& SAPA-TBF\cr
    \hline
    Cardinality         &18     &18      &20     &20\cr\hline
    Runtime (s)         &2.75   &46.272  &2.97  &52.955\cr\hline
    Matching error (dB) &-1.651 &2.628   &0.869  &3.838\cr
    \hline
    \end{tabular}
\end{table}

\section{Conclusion}
In this paper, a new sparse array beampattern design method is proposed. To prompt the sparsity of array configuration, the Shannon entropy function is imposed. However, the resulting optimization problem, which has the summation of quadratic and concave objective function, is highly nonconvex. Hence, the majorization-based ADMM algorithm is developed to solve this problem. Compared to the existing method, the proposed algorithm shows improvements in terms of beampattern matching and computational efficiency, rendering it attractive for use in radar and wireless communication systems.

\ifCLASSOPTIONcaptionsoff
  \newpage
\fi

%
%

\bibliographystyle{IEEEtran}
\bibliography{ref}

\begin{thebibliography}{10}
\providecommand{\url}[1]{#1}
\csname url@samestyle\endcsname
\providecommand{\newblock}{\relax}
\providecommand{\bibinfo}[2]{#2}
\providecommand{\BIBentrySTDinterwordspacing}{\spaceskip=0pt\relax}
\providecommand{\BIBentryALTinterwordstretchfactor}{4}
\providecommand{\BIBentryALTinterwordspacing}{\spaceskip=\fontdimen2\font plus
\BIBentryALTinterwordstretchfactor\fontdimen3\font minus
  \fontdimen4\font\relax}
\providecommand{\BIBforeignlanguage}[2]{{%
\expandafter\ifx\csname l@#1\endcsname\relax
\typeout{** WARNING: IEEEtran.bst: No hyphenation pattern has been}%
\typeout{** loaded for the language `#1'. Using the pattern for}%
\typeout{** the default language instead.}%
\else
\language=\csname l@#1\endcsname
\fi
#2}}
\providecommand{\BIBdecl}{\relax}
\BIBdecl

\bibitem{01}
H.~{He}, P.~{Stoica}, and J.~{Li}, ``Wideband \protect{MIMO} systems: Signal
  design for transmit beampattern synthesis,'' \emph{IEEE Transactions on
  Signal Processing}, vol.~59, no.~2, pp. 618--628, 2011.

\bibitem{02}
B.~{Chen}, X.~{Chen}, Y.~{Huang}, and J.~{Guan}, ``Transmit beampattern
  synthesis for the \protect{FDA} radar,'' \emph{IEEE Antennas and Wireless
  Propagation Letters}, vol.~17, no.~1, pp. 98--101, 2018.

\bibitem{03}
W.-Q. Wang and Z.~Zheng, ``Hybrid \protect{MIMO} and phased-array directional
  modulation for physical layer security in mmwave wireless communications,''
  \emph{IEEE Journal on Selected Areas in Communications}, vol.~36, no.~7, pp.
  1383--1396, 2018.

\bibitem{04}
Y.~Alsaba, S.~K.~A. Rahim, and C.~Y. Leow, ``Beamforming in wireless energy
  harvesting communications systems: A survey,'' \emph{IEEE Communications
  Surveys Tutorials}, vol.~20, no.~2, pp. 1329--1360, 2018.

\bibitem{05}
P.~{Pal} and P.~P. {Vaidyanathan}, ``Nested arrays: A novel approach to array
  processing with enhanced degrees of freedom,'' \emph{IEEE Transactions on
  Signal Processing}, vol.~58, no.~8, pp. 4167--4181, 2010.

\bibitem{06}
E.~{BouDaher}, Y.~{Jia}, F.~{Ahmad}, and M.~G. {Amin}, ``Multi-frequency
  co-prime arrays for high-resolution direction-of-arrival estimation,''
  \emph{IEEE Transactions on Signal Processing}, vol.~63, no.~14, pp.
  3797--3808, 2015.

\bibitem{07}
B.~{Fuchs}, ``Synthesis of sparse arrays with focused or shaped beampattern via
  sequential convex optimizations,'' \emph{IEEE Transactions on Antennas and
  Propagation}, vol.~60, no.~7, pp. 3499--3503, 2012.

\bibitem{08}
S.~E. {Nai}, W.~{Ser}, Z.~L. {Yu}, and H.~{Chen}, ``Beampattern synthesis for
  linear and planar arrays with antenna selection by convex optimization,''
  \emph{IEEE Transactions on Antennas and Propagation}, vol.~58, no.~12, pp.
  3923--3930, 2010.

\bibitem{09}
S.~A. {Hamza} and M.~G. {Amin}, ``Sparse array design for transmit
  beamforming,'' in \emph{2020 IEEE International Radar Conference (RADAR)},
  2020, pp. 560--565.

\bibitem{10}
J.~{Liang}, X.~{Zhang}, H.~C. {So}, and D.~{Zhou}, ``Sparse array beampattern
  synthesis via alternating direction method of multipliers,'' \emph{IEEE
  Transactions on Antennas and Propagation}, vol.~66, no.~5, pp. 2333--2345,
  2018.

\bibitem{11}
R.~C. {Nongpiur} and D.~J. {Shpak}, ``Synthesis of linear and planar arrays
  with minimum element selection,'' \emph{IEEE Transactions on Signal
  Processing}, vol.~62, no.~20, pp. 5398--5410, 2014.

\bibitem{12}
T.~{Hong}, X.~{Shi}, and X.~{Liang}, ``Synthesis of sparse linear array for
  directional modulation via convex optimization,'' \emph{IEEE Transactions on
  Antennas and Propagation}, vol.~66, no.~8, pp. 3959--3972, 2018.

\bibitem{13}
B.~{Fuchs}, ``Antenna selection for array synthesis problems,'' \emph{IEEE
  Antennas and Wireless Propagation Letters}, vol.~16, pp. 868--871, 2017.

\bibitem{14}
Z.~{Zheng}, Y.~{Fu}, and W.~Q. {Wang}, ``Sparse array beamforming design for
  coherently distributed sources,'' \emph{IEEE Transactions on Antennas and
  Propagation}, pp. 1--1, 2020.

\bibitem{15}
W.~{Fan}, J.~{Liang}, X.~Fan, and H.~C. {So}, ``A unified sparse array design
  framework for beampattern synthesis,'' \emph{Signal Processing}, vol. 182, p.
  107930, 2021.

\bibitem{16}
S.~{Huang} and T.~D. {Tran}, ``Sparse signal recovery via generalized entropy
  functions minimization,'' \emph{IEEE Transactions on Signal Processing},
  vol.~67, no.~5, pp. 1322--1337, 2019.

\bibitem{17}
P.~{Xiao}, B.~{Liao}, and J.~{Li}, ``One-bit compressive sensing via
  \protect{S}chur-concave function minimization,'' \emph{IEEE Transactions on
  Signal Processing}, vol.~67, no.~16, pp. 4139--4151, 2019.

\bibitem{18}
\BIBentryALTinterwordspacing
P.~{Xiao}, P.~{Chu}, and B.~{Liao}, ``\protect{ADMM}-based approach for
  compressive sensing with negative weights,'' \emph{IET Signal Processing},
  vol.~14, pp. 854--860(6), 2020. [Online]. Available:
  \url{https://digital-library.theiet.org/content/journals/10.1049/iet-spr.2020.0276}
\BIBentrySTDinterwordspacing

\bibitem{19}
C.~{Lu}, J.~{Feng}, S.~{Yan}, and Z.~{Lin}, ``A unified alternating direction
  method of multipliers by majorization minimization,'' \emph{IEEE Transactions
  on Pattern Analysis and Machine Intelligence}, vol.~40, no.~3, pp. 527--541,
  2018.

\bibitem{20}
Y.~Sun, P.~Babu, and D.~P. Palomar, ``Majorization-minimization algorithms in
  signal processing, communications, and machine learning,'' \emph{IEEE
  Transactions on Signal Processing}, vol.~65, no.~3, pp. 794--816, 2017.

\bibitem{21}
\BIBentryALTinterwordspacing
Z.-Q. Luo, N.~D. Sidiropoulos, P.~Tseng, and S.~Zhang, ``Approximation bounds
  for quadratic optimization with homogeneous quadratic constraints,''
  \emph{SIAM Journal on Optimization}, vol.~18, no.~1, pp. 1--28, 2007.
  [Online]. Available: \url{https://doi.org/10.1137/050642691}
\BIBentrySTDinterwordspacing

\bibitem{22}
T.~{Vu}, R.~{Raich}, and X.~{Fu}, ``On convergence of projected gradient
  descent for minimizing a large-scale quadratic over the unit sphere,'' in
  \emph{2019 IEEE 29th International Workshop on Machine Learning for Signal
  Processing (MLSP)}, 2019, pp. 1--6.

\end{thebibliography}

\ifCLASSOPTIONcaptionsoff
  \newpage
\fi

\end{document}